# Environmentally friendly method of silicon recycling: synthesis of silica nanoparticles in an aqueous solution

J.V. Bondareva[a], T.F. Aslyamov[a], A.G. Kvashnin[b], P.V. Dyakonov[a], Y.O. Kuzminova[a], Yu.A. Mankelevich[c], E.N. Voronina[c,d], S.A. Dagesyan[d], A.V. Egorov[e], R.A. Khmelnitsky[f], M.A. Tarkhov[g], N.V. Suetin[c], I.S. Akhatov[a] and S.A. Evlashin[a*]

[a]Center for Design, Manufacturing & Materials, Skolkovo Institute of Science and Technology, 30, bld. 1 Bolshoy Boulevard Moscow, 121205, Russia

[b]Center for Electrochemical Energy Storage, Skolkovo Institute of Science and Technology, 30, bld. 1 Bolshoy Boulevard Moscow, 121205, Russia

[c]Skobel'tsyn Institute of Nuclear Physics, Lomonosov Moscow State University, 1-2 Leninskiye gory, Moscow, 119991, Russia

[d]Faculty of Physics, Lomonosov Moscow State University, 1-2 Leninskiye gory, Moscow, 119991, Russia

[e]Department of Chemistry, Lomonosov Moscow State University, 1-3 Leninskiye Gory, Moscow, 119991, Russia

[f]P. N. Lebedev Physical Institute, Russian Academy of Sciences, 53 Leninsky Prospect, Moscow, 119991, Russia

[g]Institute of Nanotechnology of Microelectronics of the Russian Academy of Sciences, Leninsky Prospect, 32A, Moscow, 119991, Russia

*e-mail: s.evlashin@skoltech.ru

## Abstract

Future decades will experience tons of silicon waste from various sources, with no reliable recycling route. The transformation of bulk silicon into $SiO_2$ nanoparticles is environmentally significant because it provides a way to recycle residual silicon waste. To address the needs of silicon recycling, we develop a top-down approach that achieves 100% conversion of bulk silicon to silica nanoparticles with outcome sizes of 8-50 nm. In addition to upcycling the potential of silica, our method also possesses several advantages, such as simplicity, scalability and controllable particle size distribution. Many fields of science and manufacturing, such as optics, photonics, medical, and mechanical applications, require size-controllable fabrication of silica nanoparticles. We demonstrate that control over temperature and hydrolysis time has a significant impact on the average particle size and distribution shape. Additionally, we unravel the process of nanoparticle formation using a theoretical nucleation model and quantum density functional theory calculations. Our results provide a theoretical and experimental basis for silica nanoparticle fabrication and pave the way for further silicon conservation research.

## Introduction

The choice of silicon for the synthesis of silicon dioxide nanoparticles is determined by its availability and wide range of applications in electronics for the manufacture of semiconductor materials and photovoltaics for the production of solar cells. The majority of photovoltaic (PV) cells are made from silicon, and their production is increasing year by year[1]. The durability of solar panels is usually not more than a few decades. Over their lifetime, the panels slowly degrade and produce less electricity. If recycling does not occur, there will be 60 million tons of PV panel waste lying in landfills by the year 2050[2]. The problem of solar panel disposal will manifest itself in two or three decades and will considerably challenge the environment due to the vast amounts of waste, recycling of which is not economically attractive. The scientific community is paying increasing attention every year to environmental impacts. One promising route of reducing environmental damage from solar panel waste is etching away and recycling the silicon wafers. An effective way to convert bulk silicon plates into silica nanoparticles can recycle used-up silicon for further application in material treatment, optics, photonics, and medical applications.

Silica nanoparticles are emerging nanomaterials due to their chemical and thermal stability, safety, optical transparency, and commercial availability[3]. Synthetic methods of fabrication allow the production of particles with a high degree of purity and distinct size distribution compared to natural sources[4]. The synthesis of spherical silica nanoparticles is of great interest due to their potential application in medicine for the encapsulation of enzymes[5], drug delivery systems[6], and drug release[7], and in the industry for the production of electronic devices[8], nanocomposites, efficient adsorbents[9], and catalysts[10-18]. In a recent publication[19], even a low concentration of silica nanoparticles (<50 nm) crucially influenced the rheological properties of the surfactant. The properties of nanoparticles, which determine their further application, strongly depend on the synthesis conditions.

Generally, approaches for the synthesis of nanostructures can be divided into two groups: top-down and bottom-up[20, 21]. There are examples of methods for producing silicon dioxide nanoparticles using the Stober bottom-up approach or by creating a microemulsion in which alkoxy silicates (TEOS, TMOS) or inorganic salts (sodium silicate) are used as silicon sources[4]. The preparation of spherical particles of silicon dioxide by hydrolysis of tetraethoxysilane (TEOS) in an alcohol solution in the presence of ammonia as a catalyst was first described by Kolbe in 1956[22]. In 1968, Stober and coauthors improved this method to produce monodisperse $SiO_2$ particles with diameters ranging from tens of nanometers to several micrometers[23-25]. The Stober multistep method is based on the hydrolysis of silicon alkoxides in a water-alcohol medium in the presence of ammonium hydroxide as a catalyst. It allows the production of particles with controllable size[26]. Silicon alkoxides can be hydrolyzed under both basic and acidic conditions that lead to polycondensation and the formation of siloxane bonds (Si – O – Si). As a result of this procedure, a sol of dispersed nanoparticles forms.

The main challenges of the Stober method are the inability to produce particles with a diameter of less than 100 nm with a narrow size distribution, the usage of large amounts of ethanol and a long synthesis duration. The Strober method was modified to reduce the particle size to 40 nm by changing the synthesis conditions: ammonia concentration, reaction temperature, and type of solvent[27]. The possibility of producing 12–23 nm spherical $SiO_2$ particles by hydrolysis of tetraethoxysilane in a cyclohexane solution in the presence of amino acids was demonstrated[28]. The use of amino acids allowed control of the particle size and generation of particles with a diameter of less than 50 nm with a high degree of monodispersity.

An alternative to Strober's approach for silica nanoparticle synthesis is the microemulsion method[29,30]. In this approach, surfactant molecules are dissolved in organic solvents to form spherical micelles. In an aqueous medium, polar groups form cavities in which the synthesis of nanoparticles occurs at a specific ratio of silicon precursor and catalyst. The main disadvantages of the microemulsion method are the high cost and difficulties associated with purification of the final product from a surfactant.

In this study, we propose a simple one-step transformation process of bulk silicon to silica nanoparticles with 100% efficiency. Silica nanoparticles were obtained by dissolution of bulk silicon in aqueous solutions with pH values higher than five. Variation in the temperature and time of the synthesis allows control of the mean size of the generated particles in the range from 8 to 50 nm. Universal Structure Predictor: Evolutionary Xtallography (USPEX) was used to explain the etching mechanism of silicon in a water medium. Since our approach does not involve any toxic reagents and allows us to convert bulk silicon, it is of great environmental significance in areas of science and production dealing with waste silicon.

## Results

Silica nanoparticles can be obtained by the simple hydrothermal decomposition of bulk silicon in pure water under autogenous pressure. SEM images of the nanoparticles obtained using pure water are shown in Fig. 1a. Since this process proceeds rather slowly, a base was added to the system as a catalyst to accelerate the reaction rate. In the literature, to obtain silica nanoparticles by the Stober method, an aqueous solution of ammonia is used as a base[29]. This base is used to achieve the optimal reaction rate, which allows us to reduce the synthesis time and conduct a detailed study of the reaction products. At the same time, an aqueous solution of ammonia is not the only base that can be used for the synthesis of silica nanoparticles. We experimentally studied a large number of bases, and the results of this research are shown in Fig. 1b. We can obtain silica nanoparticles using aqueous solutions of several organic and inorganic bases with a pH value higher than nine and aqueous solutions of inorganic salts, giving a slightly alkaline reaction. Considering that the synthesis of silica nanoparticles proceeds in an aqueous medium in the absence of any bases, it can be concluded that the reaction occurs at a pH value higher than five.

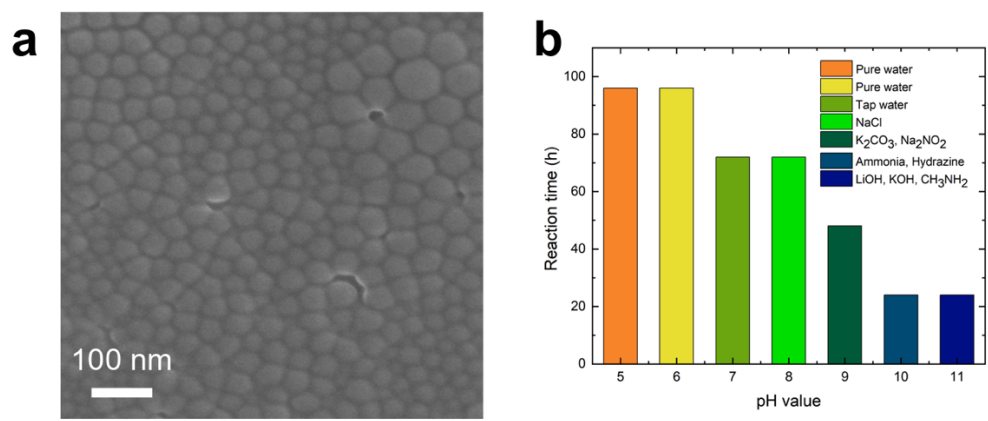

Figure 1: Silicon nanoparticles reaction media. (a) SEM image of silicon nanoparticles produced in pure water during one week. (b) Dependencies of the reaction time (time for dissolving 0.3 g silicon) on the pH value.

To reduce the synthesis time, all of the silica samples described below were obtained by adding aqueous ammonia to the reaction mixture. Fig. 2 shows the SEM images of spherical $SiO_2$ nanoparticles with different synthesis times (48 and 96 h); the amount of silicon was 0.3 g, and the temperature was 453.15 K. Fig. 2a depicts spherical $SiO_2$ nanoparticles synthesized for 48 h. Using SEM images of the particles obtained via these experimental conditions, a particle size distribution (PSD) close to a normal distribution was obtained. However, an increase in the process time to 96 h leads to an increase in the mean particle size, as can be observed in the SEM images (Fig. 2b). For these synthesis parameters, PSD possesses a bimodal shape, which is revealed to be the sum of two normal distributions. More information on the PSD and SEM images can be found in the Supporting Information (Fig. S1, Fig. S2, Fig. S3, Fig. S4, Fig. S5). Overall, an increase in reaction time leads to an increase in the average particle size and a bimodal PSD shape.

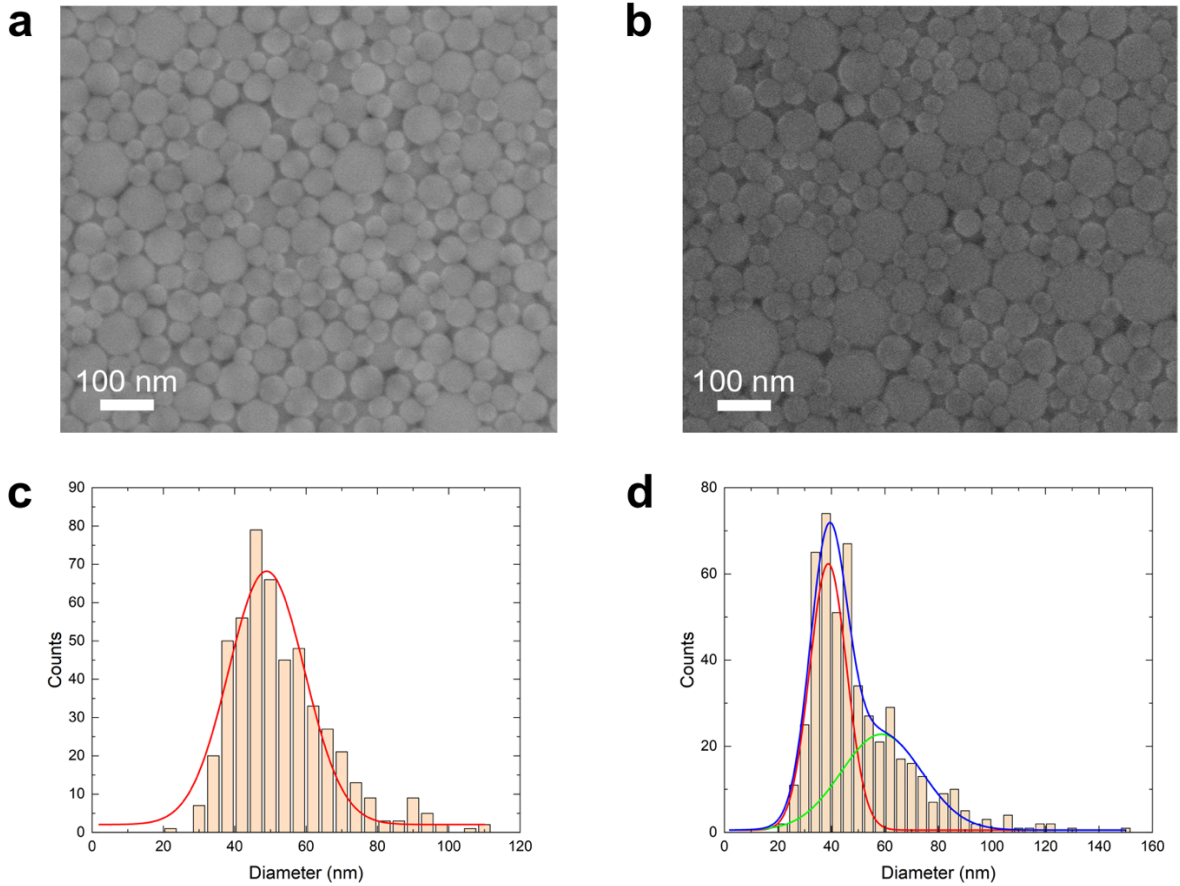

Figure 2: SEM images of spherical SiO$_2$ nanoparticles and PSD histograms produced at 453.15 K. (a), (c) Reaction time is 48 h. (b), (d) Reaction time is 96 h.

For a more accurate visualization of the obtained particles, the studies were supplemented by observations under a transmission electron microscope. The combination of these techniques provides us with visualization of individual nanoparticles, associated particle-agglomerate morphologies and a detailed internal structure of the particles. Fig. 3 shows that the prepared silica particles are spherical and amorphous and are approximately equal in size.

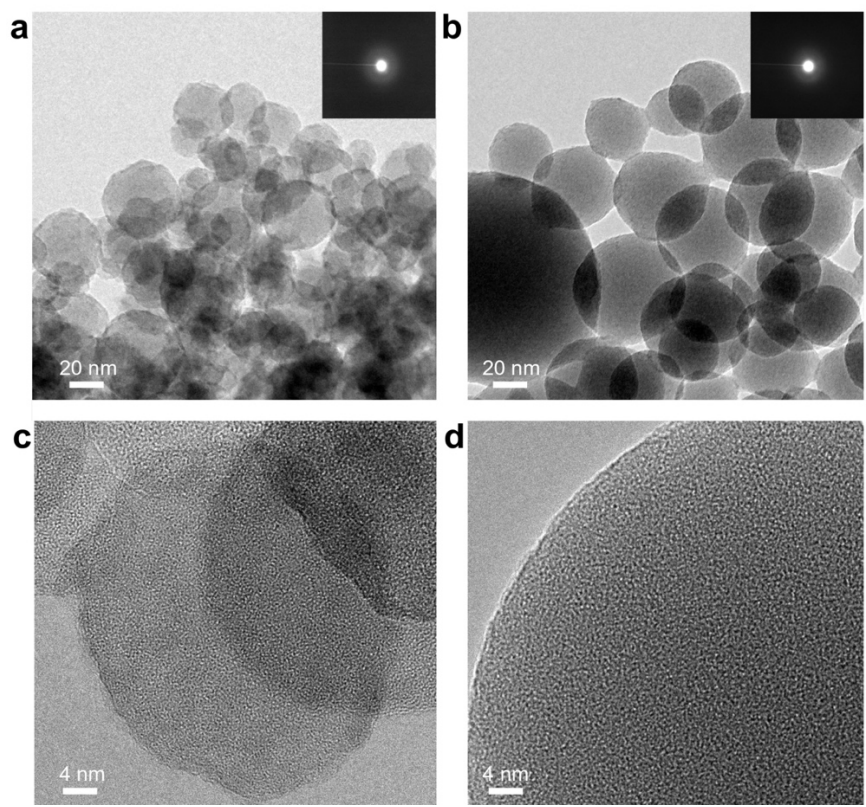

Figure 3: TEM images of silica nanoparticles synthesized at different temperatures. (a) Synthesis temperature is 353.15 K. (b) Synthesis temperature is 453.15 K. The inset shows the corresponding highly diffusive ring pattern of electron diffraction. (c), (d) High-resolution TEM micrographs of silica nanoparticles synthesized at 353.15 K and 453.15 K.

The highly diffusive ring pattern in the corresponding selected area electron diffraction shown in the inset reveals that the obtained silica is in a completely amorphous state. From the TEM images, it can also be seen that silica nanoparticles tend to aggregate together. Fig. 3 c, d shows the internal structure of silica nanoparticles. X-ray diffraction also shows that the particles are amorphous (see Fig. S6).

We have shown that FTIR and NMR techniques can be easily employed to monitor the conversion of bulk silicon into insoluble $SiO_2$. The FTIR spectrum in the 1500-450 cm$^{-1}$ range of the obtained silica nanoparticles deposited on the film and dried in air and at 453.15 K are shown in Fig. S7. The three main peaks characteristic of Si-O-Si vibrational modes[31] are detected at approximately 1070, 800, and 460 cm$^{-1}$. The FTIR spectrum of the nanoparticles shows three main characteristic peaks attributed to the Si-O-Si bonds, which are assigned to the asymmetric vibration of Si–O (1108 cm$^{–1}$), the symmetric vibration of Si–O (815 cm$^{–1}$) and rocking motions of Si–O (460 cm$^{–1}$). The MAS NMR spectrum of $^{29}$Si nuclei contains three characteristic signals related to silicon atoms in a tetrahedral oxygen environment (Q2) (-93.9 ppm) with silanediol groups (Q3) (-101.4 ppm) and a silanol group but without silanol groups (Q4) (-111.9 ppm), as presented in Fig. S7. The high signal intensity from the centers Q2 and Q3 indicates that the material has a large number of silanol groups (mainly Q3) on the surface[31,32].

The adsorption properties of two samples of silica materials, which were synthesized for 48 h at temperatures of 353 K and 453 K, were studied. Analysis of the adsorption isotherms shows that the specific surface area (SSA) of the samples prepared at 353 and 453 K is equal to 45.9 m$^2$/g and 10.9 m$^2$/g, respectively. With increasing synthesis temperature, the SSA significantly decreases. This behavior and the numerical SSA values are consistent with similar measurements for popular silica materials such as CPG and LiChrospher[33]. (For more information, see the Supplementary Information).

## Discussion

An evolutionary search for the thermodynamically stable adsorption structure of water on the Si(100)-(2×1) surface allowed us to find five structures, as shown in Fig. 4c. We denote the obtained structures as n·H$_2$O, where n is the number of water molecules added on the surface of the considered unit cell of Si(100)-(2×1) having a surface area of 58.4 Å$^2$. One can see that in all cases, the water molecules split into oxygen and hydrogen to be adsorbed on the Si(100)-(2×1) surface. Due to the reconstruction of the Si(100) surface, the outer Si atoms have dangling bonds responsible for the chemical activity. Thus, it is energetically more favorable to chemisorb oxygen or hydrogen on the surface rather than to physisorb them. However, in the case of a high concentration of $H_2O$ (5 molecules per unit cell), the physisorption of $H_2O$, as well as $H_2$ molecules, was observed (Fig. 4a, 5·$H_2O$). In the case of 4·$H_2O$, there is one $H_2$ molecule that stays unbonded to the surface (the distance from the surface is ~5 Å; not shown in Fig. 4a).

The calculated dependence of the surface energy on the chemical potential of water is shown in  Fig. 5b. Here, $\Delta\mu_{H2O}$ is the difference between the chemical potentials of water in solid and gas states ($\mu_{H_2O}^{gas} = -14.22\ eV, \mu_{H_2O}^{ice_{Ih}} = -14.92\ eV$). The ice structure in the Ih modification was used to calculate the upper bound of the water chemical potential. The surface energy of pure Si(100)-(2×1) was found to be 80.02 meV/Å$^2$, which agrees well with data from[34]. It should be noted that the addition of water molecules on the surface significantly reduces the surface energy (Fig. 4b). One can see that the adsorption of a single water molecule leads to a reduction in the surface energy to 25 meV/Å$^2$ ($\Delta\mu_{H_2O}= 0$). A decrease in the chemical potential difference (heating of water) leads to a decrease in the surface energy and to the formation of more complex surface structures with more water molecules adsorbed. Thus, at a chemical potential difference of -0.31 eV, the 2·$H_2O$ structure becomes stable (blue color in Fig. 4b). Further changes in the water chemical potential lead to subsequent stabilization of the adsorption structures with n = 3, 4 and 5. The 3·$H_2O$ surface structure has a very narrow stability range of $\Delta\mu_{H_2O}$ between -0.46 and -0.47 eV (see inset of Fig. 1b). The 4·$H_2O$ surface structure is stable from $\Delta\mu_{H_2O} = -0.47$ eV to -0.65 eV (green line in Fig. 4b). The surface structure 5·$H_2O$ is stable from $\Delta\mu_{H_2O}= -0.65$ eV to -0.7 eV (the lower bound of the chemical potential range corresponding to the chemical potential of water molecules in the gas phase).

The calculated pressure-temperature phase diagram of the predicted surface structure shown in Fig. 4c makes it possible to predict the environmental conditions (partial water pressure and temperature) suitable for the formation of a particular surface structure. Both partial water pressure and temperature are included in the expression for chemical potential:

$$\mu_{H_2O} = E_{H_2O} + G_{H_2O}(T, P_0) + k_B T \ln\left(\frac{P}{P_0}\right) = E_{H_2O} + \Delta\mu_{H_2O}(T, P), \quad (1)$$

where $G_{H_2O}(T, P_0)$ is the calculated Gibbs free energy of the gas of water molecules at a certain temperature and pressure, which agrees well with the data from the thermodynamic database[35].

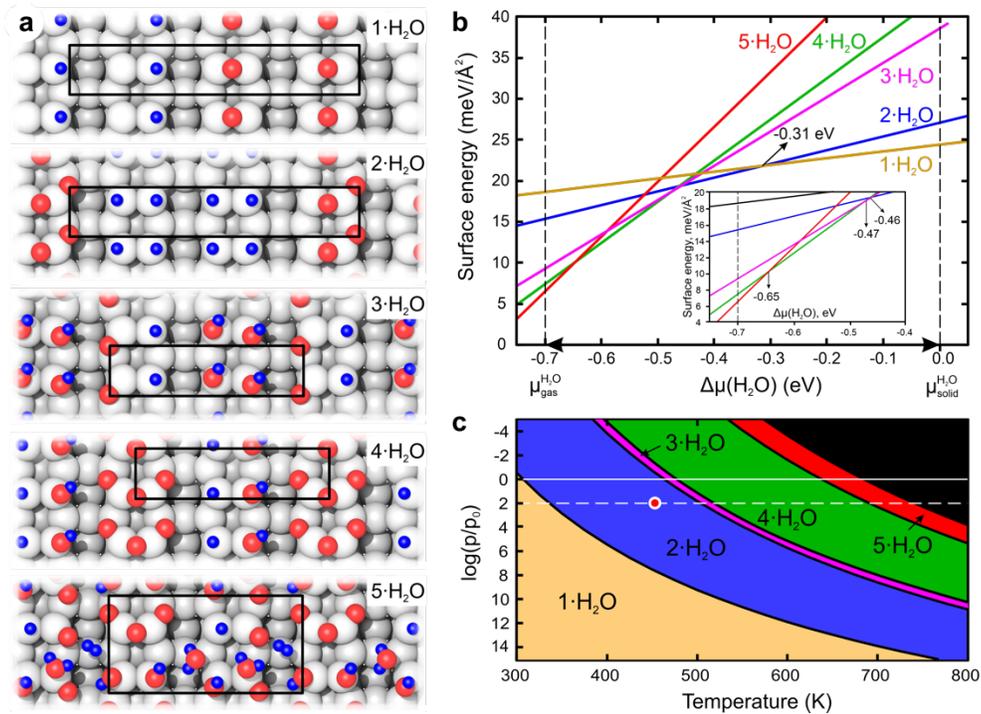

Figure 4: Calculated surface energies of predicted reconstructions. (a) Atomic structures of energetically favorable adsorption structures of water on the Si(100)-(2×1) surface. (b) Surface energy as a function of $\mu_{H_2O}$ (c) Phase diagram of the Si(100)-(2×1) surface with adsorbed water molecules. The condition of our experiment is shown by a red point, corresponding to a pressure of 100 bar and a temperature of 453 K.

The red point in Fig. 4c shows the experimentally used conditions in which the silicon sample is located. According to our calculations, we propose that in the experiment, the concentration of water molecules adsorbed on the Si(100) surface at set conditions (453 K, 100 bar) was approximately $10^{18}$ molecules per square meter. We can also suggest that more water molecules will adsorb as the temperature increases, which will increase the reaction rate, and the silicon sample will be destroyed faster.

Computer simulations demonstrate the presence of SiH surface bonds under the synthesis conditions. This atomic configuration is popular in studies of the wet oxidation of silicon crystals initially acid etched (hydrogen-terminated)[36]. Therefore, the H-bonding interaction of silicon surface with the water molecules is described by the well-known formula[37]:

$$SiH + OH^- + H^+ \rightarrow SiOH + H_2, \qquad (2)$$

Assuming a sufficient concentration of $OH^-$, which is defined by pH, the formation of OH bonds may remove Si monomers ($SiH_x(OH)_y$, where x+y=4) from the surface. For example, the case of $Si(OH)_4$ was shown in[37]. In accordance with the TEOS procedure, these Si monomers undergo self-assembly to form silica nanoparticles[38]. Therefore, the pH crucially influences the rate of nanoparticle formation, in agreement with our observations (Fig. 1b).

Additionally, impurities (P and B dopants) and molecular defects on the silicon surface lead to reaction acceleration. In the case of pure water, surface modifications would shift the chemical potential to the solid phase. As one can see from Fig. 4c, it corresponds to the $3 \cdot H_2O$, $4 \cdot H_2O$, and $5 \cdot H_2O$ regions. These states exhibit already formed Si-(OH) bonds that enhance Si monomer release. In summary, these results indicate that pH and impurities influence the silicon etching process.

The reaction time affects the size of the particles. The numerical dependence of the average particle size on the reaction time is shown in Fig. 5a. The average particle size initially increases sharply and then remains almost constant. Therefore, the process of particle formation exhibits a characteristic saturation time of approximately 24 h. We considered the nanoparticles from this saturation region (with a reaction time of 24 h or more) at various temperatures in the range of 323 K to 453 K. The particle size analysis from the SEM measurements (see Supporting Information Fig. **S1**) demonstrates that the average particle size increases monotonically as the temperature increases (Fig. 5b). This experimental result coincides with the Arrhenius law, as follows:

$$d \sim exp\left(-E_a/T\right) \qquad (3)$$

where $E_a$ is the activation energy, which in our work equals $E_a = 1920K$.

The formation of nanoparticles can be described as the heterogeneous nucleation process, which may proceed at a finite number of nucleation sites distributed on the surface of silicon. We analyzed the experimental time dependence in Fig. 5a using the phase transformation Avrami (also known as Johnson-Mehl-Avrami-Kolmogorov) equation:

$$(d/d_0)^3 = 1 - exp(-k\, t^n), \qquad (4)$$

where $d$ is the measured average particle diameter and $d_0$ is the characteristic size; the ratio $d/d_0$ relates the degree of the phase transition, which equals 0 and 1 in the initial and new phases, respectively; and k and n are the model constants. According to the original Avrami theory, n is the integer number between 1 and 4 describing the dimension of the process. However, it was demonstrated that considering the constant number of nucleation sites on the surface, their saturation leads to n=1[39]. Additionally, in ref[40], the authors demonstrated that in the case of heterogeneous nucleation in three-dimensional space, the Avrami equation also has the parameter n=1.

The effect of the concentration of ammonia was analyzed. A decrease in the amount of ammonia does not have a substantial impact on the average particle size; for five different concentration values, it remained at approximately 40 nm, as shown in Fig. S8. With a decrease in the amount of ammonia, the particle size distribution becomes narrow; however, there are no bands in the region from 80 to 140 nm, which indicates a high degree of monodispersity of the obtained silicon dioxide. Additionally, we analyzed the effect of autoclave loading on the average particle size. The filling volume influences the pressure inside the vessel[41]. The volume of the load/pressure does affect the average particle size; for three different volume values, it remained at approximately 30 nm, as shown in Fig. S8. Therefore, the temperature and reaction time only have a notable influence on the characteristics of the synthesized nanoparticles. Thus, choosing the optimal temperature and reaction time values allows synthesis of the smallest, homogeneous, and monodisperse silica nanoparticles.

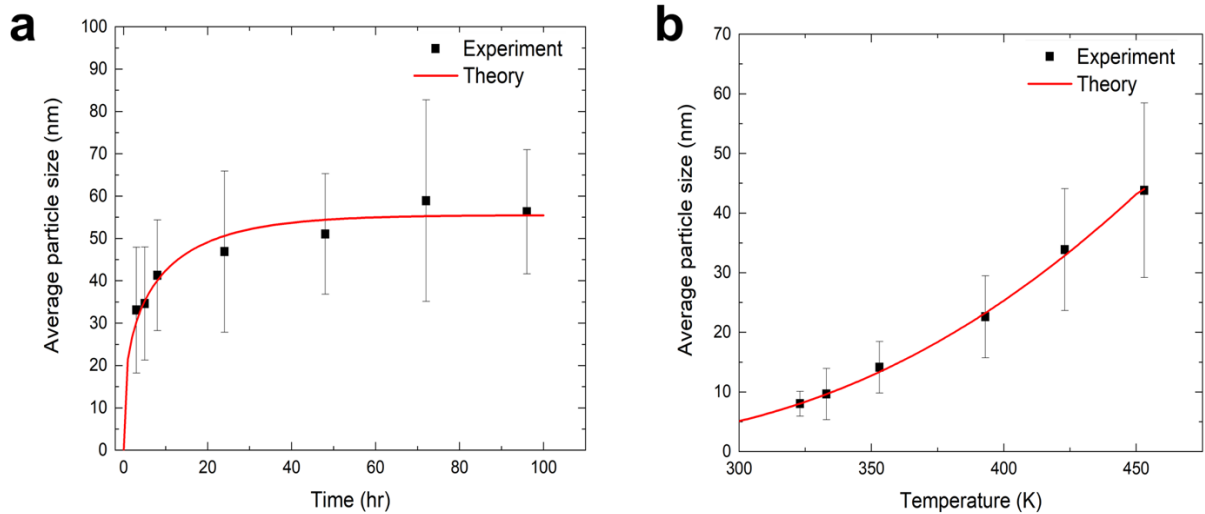

Figure 5: The dependence of the average particle size on the synthesis parameters. (a) The dependences of average particle size on the time of hydrolysis (T = 453.15 K). (b) The dependences of average particle size on temperature ($C_{ammonia}$ = const = 1%).

## Conclusions

We have developed a hydrothermal top-down approach for silica nanoparticle synthesis. Our approach recycles bulk silicon sealed in an autoclave filled with aqueous ammonia medium at various temperatures and reaction times. We characterized the synthesized nanoparticles using SEM, TEM and adsorption measurements. Our analysis shows that the temperature and reaction time have a crucial impact on the particle size distribution and adsorption characteristics. More precisely, the average size increases as the temperature or reaction time increases. We synthesized particles with controllable sizes in the range of 8-50 nm at temperatures from 300 K to 453 K and used reaction times from 2 h to 96 h. The process of particle formation is described by the heterogeneous nucleation model, which fits the experimental data well. Moreover, silica nanoparticles can be obtained in an aqueous medium in the absence of bases. The etching mechanism of silicon in an aqueous medium was explained using ab initio calculations. The described synthesis method possesses considerable environmental significance due to its scalability, lack of toxic reagents and potential to recycle waste silicon wafers. In addition, experimental data and consistent theoretical models lay the foundation for further research on the silica conversation process.

## Materials and Methods

### Reagents

HR-, N- and P-type silicon wafers with electrical resistivities of 10 k$\Omega$ ×cm, 4.5 and 12 $\Omega$ ×cm were used as silicon sources. Ammonium hydroxide (25% aqueous solution, SigmaTek) was used without any further purification. Deionized water ($\geq$ 18 M$\Omega$ cm$^{-1}$) was used as the reaction medium. Organic bases (methylamine, triethylamine, hydrazine, tetramethylammonium hydroxide), inorganic bases (LiOH, NaOH, KOH), inorganic salts ($K_2CO_3$, $Na_2NO_2$, NaCl, $NaHCO_3$), and acids (hydrochloric acid and hydrofluoric acid) were purchased from Ekos-1 and used as alternative media.

### Synthesis of silica nanoparticles

To synthesize silica nanoparticles, pure water was used. An aqueous solution of ammonium was used to accelerate the process of particle production. Silicon was added to 60 ml aqueous ammonia medium with concentrations varying from 0.53 M to 13.16 M. These solutions were transferred into Teflon lined sealed stainless steel autoclaves and maintained at various temperatures in the range of 297.15 K to 453.15 K for times varying from 2 h to 96 h under autogenous pressure. It was then allowed to cool naturally to room temperature. After the reaction was completed, the resulting white solid products were dried in air in a laboratory oven at 453.15 K. The inner autoclave pressure varied depending on the autoclave filling of 57, 47, 37, and 27 ml.

The same procedure was carried out for the synthesis of silica nanoparticles in the presence of the alternative organic and inorganic bases and salts mentioned above with concentrations varying from 0.1 M to 1 M. The synthesis of silica nanoparticles was also successfully carried out in an aqueous medium without the addition of bases and salts as a catalyst.

Characterization

The morphology and dimensions of silica particles were studied using scanning electron microscopy (SEM) in a Carl Zeiss Supra 40 system and transmission electron microscopy (TEM) in a JEM 2100F (UHR/Cs) system with an acceleration voltage of 200 kV.

To obtain statistics on particle sizes, SEM images of the particles in each process were analyzed with LabView software, which allows us to accumulate statistics on particle sizes for several images. The sizes of as many as 500 particles were used to form a distribution and estimate the standard deviation for each process.

Fourier transform infrared (FTIR) absorption spectra, in the range 1500-400 cm$^{-1}$, were recorded using a Bruker Vertex 70 V spectrophotometer. The spectra were obtained via transmission from films deposited on a silicon substrate, which is IR transparent, with an accuracy of 0.5 cm$^{-1}$.

Crystalline structures were analyzed with a Huber G670 diffractometer with CoKα radiation (wavelength (WL) = 1.78892 Å).

Solid-state $^{29}$Si CP/MAS NMR spectra were recorded on a BRUKER AVANCE-II 400 WB spectrometer at frequencies of 400.13 and 79.5 MHz using a 4 mm H/X MAS WVT sensor. The powder samples were placed in a pencil-type zirconia rotor with a diameter of 4.0 mm. The spectra were obtained at a spinning speed of 8 kHz. The Si signal of tetramethylsilane (TMS) at 0 ppm was used as the reference for the $^{29}$Si chemical shift.

The adsorption-desorption isotherms of nitrogen were measured by an ASAP2020 volumetric adsorption analyzer (Micromeritics) at 77 K. We used the BET method to calculate SSA from experimental adsorption isotherms in the low-pressure range $0.05 \leq P/P_0 \leq 0.1$, where $P_0$ is the nitrogen saturation pressure at 77 K.

Calculations

Stable configurations of H$_2$O molecules on the Si(100) surface were predicted using the first-principles evolutionary algorithm USPEX[42,43] adopted for surfaces[44]. During the evolutionary search, we included all the surface supercells up to index 4. The unit cell of the Si(100) substrate consists of 36 atoms, while the number of H$_2$O molecules per unit cell varies from 1 to 6. Evolutionary search was combined with structure relaxations using density functional theory (DFT)[45,46] within the generalized gradient approximation (Perdew-Burke-Ernzerhof functional)[47] and projector-augmented wave method[48,49], as implemented in the VASP package[50,51,52]. The plane-wave energy cutoff was set to 400 eV, and a k-mesh of 2π×0.05 Å$^{-1}$ resolution was used. Monopole, dipole, and quadrupole corrections were taken into account using the method discussed in ref.[53,54]. The first generation of structures was produced randomly. In contrast, subsequent generations were obtained by applying 40% heredity, 10% softmutation, and 20% transmutation operations, with 30% of each new generation produced using the random symmetric algorithm[55].

Each of the supercells under study contained a vacuum layer with a thickness of 20 Å, and the thickness of the substrate was 3 unit cells of bulk Si with the topmost 4-Å layer allowed to relax. The boundary values of the physically allowed chemical potentials, which are related to the free energies of water in the gas and solid (ice$_{Ih}$) phases. The surface energies of the predicted reconstructions were calculated as:

$$\gamma = \frac{1}{N}[E_{tot} - E_{ref} - \sum_i \eta_i \mu_i] \qquad (5)$$

where $E_{tot}$ is the total energy of the whole system, $E_{ref}$ is the reference energy of the substrate (the reconstructed Si(100) surface), $\eta_i$ is the number of additional water molecules on the substrate, $\mu_i$ is the chemical potential of the water molecules, and N = m×n for an m×n surface supercell (serves as a normalization factor).

The lower limit of the chemical potential was set as the chemical potential of the gas phase of $H_2O$, while the upper limit corresponds to the case of solid water (ice$_{Ih}$). The following relation defines the physically meaningful range of chemical potentials:

$$E_{gas} \leq \mu_{H_2O}(P,T) \leq E_{ice\ ih}, \qquad (6)$$

## Author contributions

J.V.B. conducted experiments and data processing; prepared and formatted the body of the manuscript. T.F.A. was developing the scheme of experimental procedures, participated in developing of theoretical model, processed experimental data. A.G.K. - developed USPEX calculation method and executed it. P.V.D. - developed and formatted the body of the text, conducted experiments. Yu. O. K. - provided XRD analysis of the samples. Yu. A. M. - developed DFT model of the material and executed calculations. E. N. V. - developed DFT model of the material and executed calculations. S. A. D. - provided XPS analysis of the samples, A.V.E. - provided TEM analysis of the samples. R. A. Kh. - provided FTIR analysis of the samples. M.A.T. - provided DLS analysis of the samples, executed statistical analysis of the particle size. N.V.S. - performed attraction of funding, prepared and formatted the body of the manuscript. I.Sh.A. - supervised the research, developed and formatted the manuscript. S.A.E. - conducted experiment, developed and supervised the research, prepared and formatted the body of the manuscript.

Competing interests

The authors declare no competing interests.

Supporting Information for

# Environmentally friendly method of silicon recycling: synthesis of silica nanoparticles in an aqueous solution

J.V. Bondareva[a], T.F. Aslyamov[a], A.G. Kvashnin[b], P.V. Dyakonov[a], Y.O. Kuzminova[a], Yu.A. Mankelevich[c], E.N. Voronina[c,d], S.A. Dagesyan[d], A.V. Egorov[e], R.A. Khmelnitsky[f], M.A. Tarkhov[g], N.V. Suetin[c], I.S. Akhatov[a] and S.A. Evlashin[a]

[a]Center for Design, Manufacturing & Materials, Skolkovo Institute of Science and Technology, 30, bld. 1 Bolshoy Boulevard Moscow, 121205, Russia

[b]Center for Electrochemical Energy Storage, Skolkovo Institute of Science and Technology, 30, bld. 1 Bolshoy Boulevard Moscow, 121205, Russia

[c]Skobel'tsyn Institute of Nuclear Physics, Lomonosov Moscow State University, 1-2 Leninskiye gory, Moscow, 119991, Russia

[d]Faculty of Physics, Lomonosov Moscow State University, 1-2 Leninskiye gory, Moscow, 119991, Russia

[e]Department of Chemistry, Lomonosov Moscow State University, 1-3 Leninskiye Gory, Moscow, 119991, Russia

[f]P. N. Lebedev Physical Institute, Russian Academy of Sciences, 53 Leninsky Prospect, Moscow, 119991, Russia

[g]Institute of Nanotechnology of Microelectronics of the Russian Academy of Sciences, Leninsky Prospect, 32A, Moscow, 119991, Russia

*e-mail: s.evlashin@skoltech.ru

**Results**

*I. SEM*

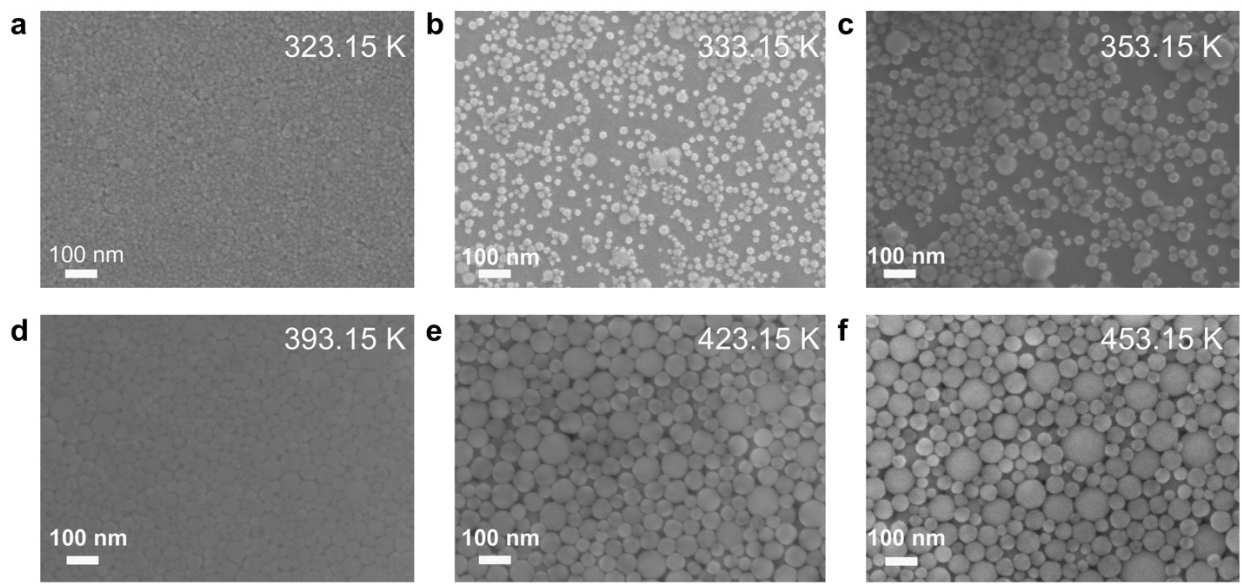

Figure S1: SEM images of silica nanoparticles at different temperatures in ammonia media.

*II. PSD*

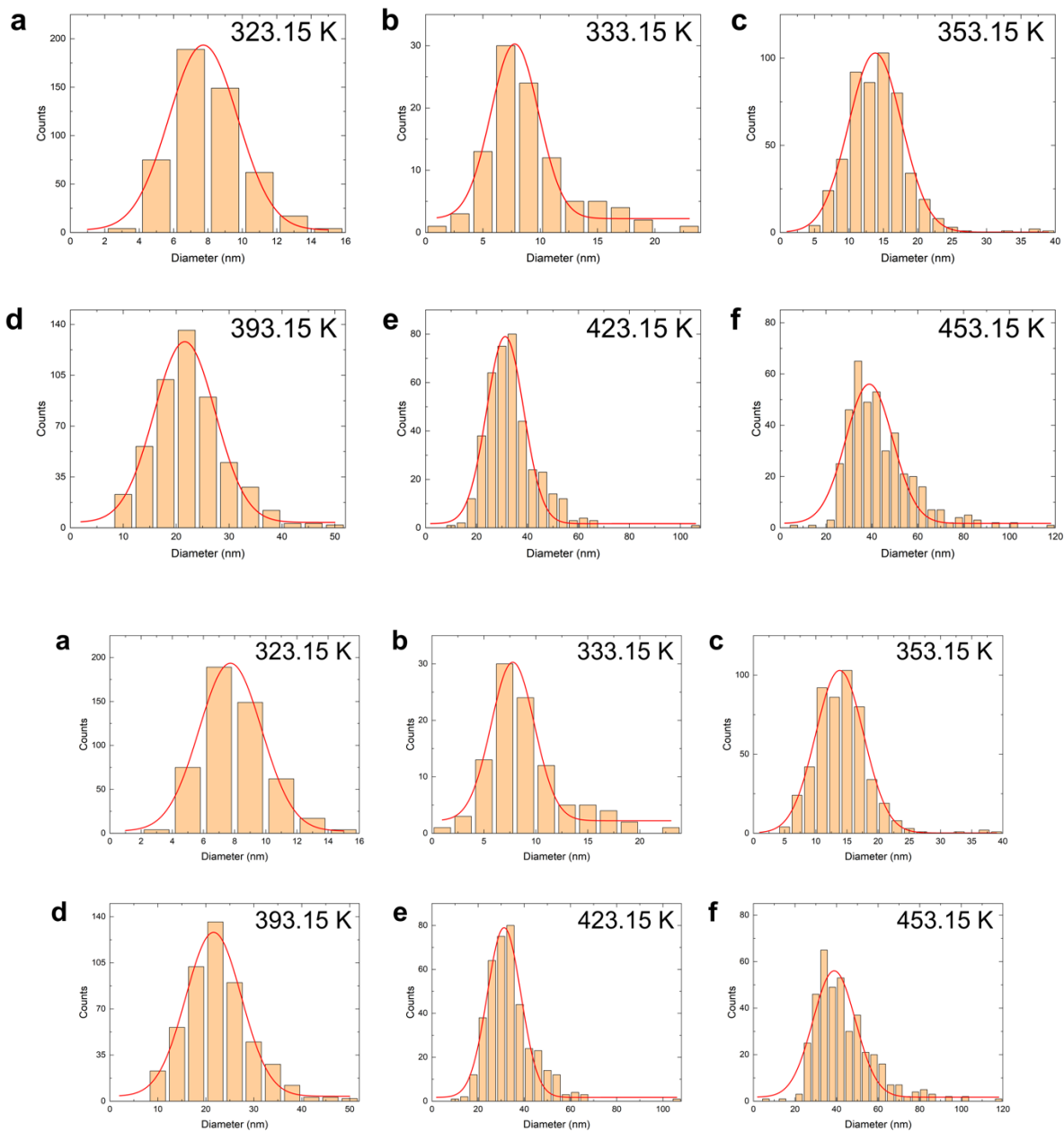

Figure S2: PSD histogram of silica nanoparticles at various temperatures in ammonia media.

*III.    SEM*

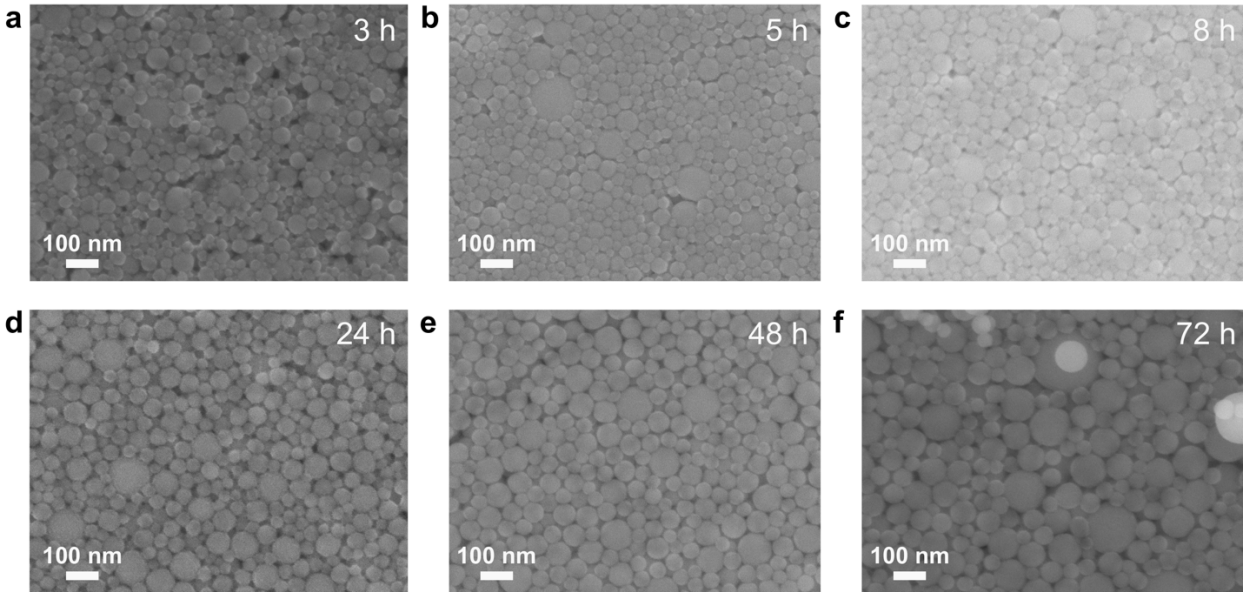

Figure S3: SEM images of silica nanoparticles at various reaction times in ammonia media.

## IV. PSD

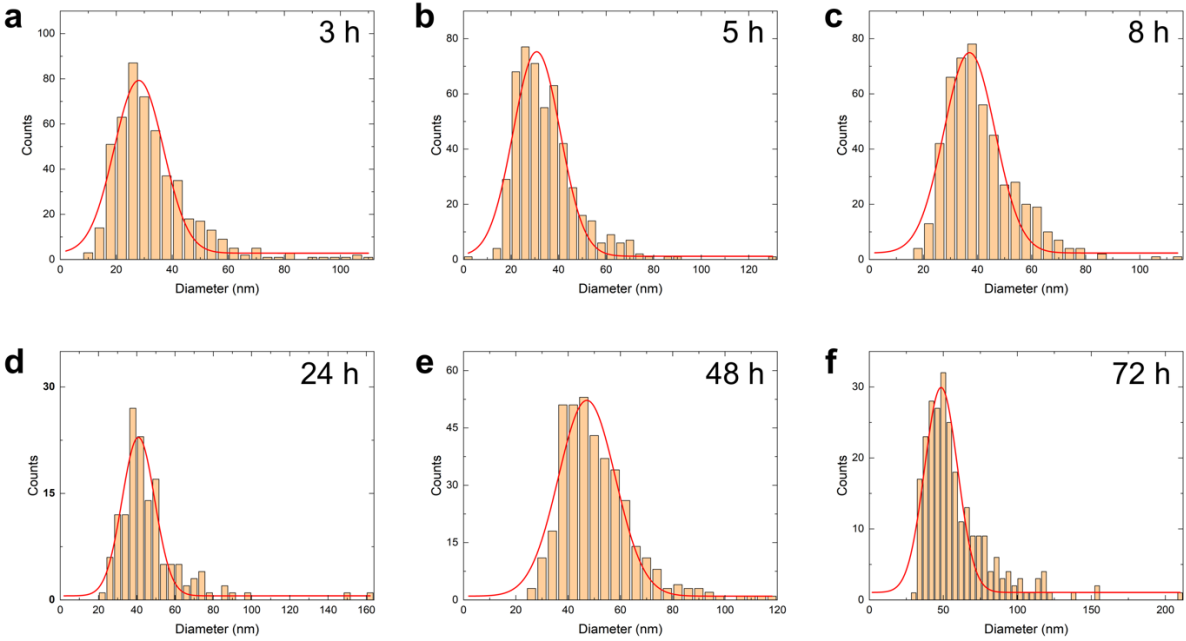

Figure S4: PSD histogram of silica nanoparticles at various reaction times in ammonia media.

## V. PSD vs concentration of ammonia

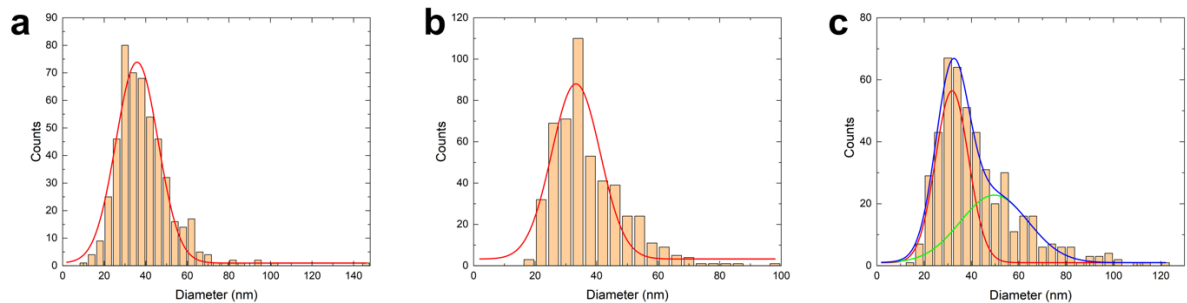

Figure S5: PSD histograms of the obtained silica nanoparticles at different amounts of ammonia. a, 6.25% b, 12.5% c, 25%.

## VI. XRD

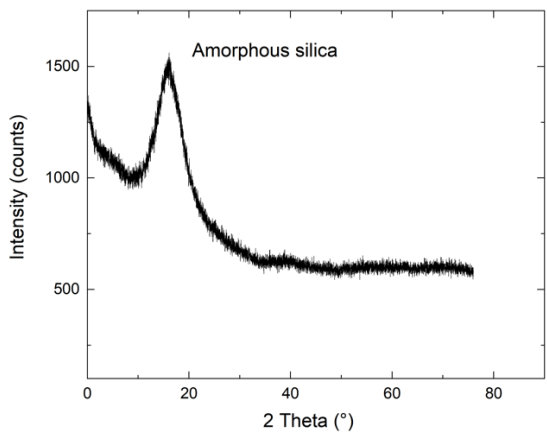

Figure S6: The X-ray diffraction pattern of the dried extracted solid indicates a broad peak at 2θ = 19°, which reveals the amorphous nature of the silica nanoparticles.

## VII. FTIR, NMR analysis

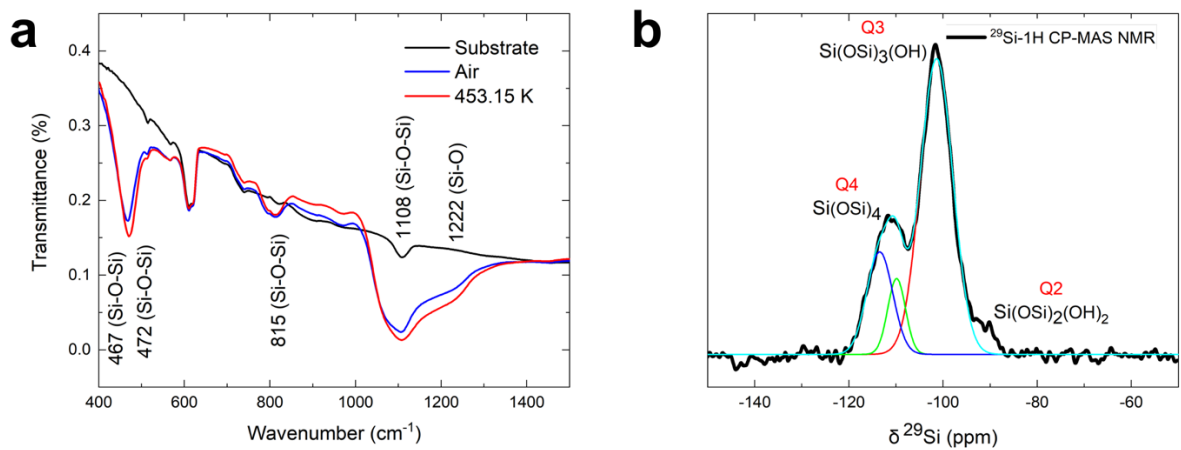

Figure S7 (a) FTIR spectra of silica nanoparticles. (b) $^{29}$Si CP/MAS NMR spectra of silica nanoparticles.

## VIII.   *Absorption*

The SEM measurements demonstrate the existence of nanoscale pores between synthesized silica particles. Additionally, the increase in the synthesis temperature leads to the formation of relatively large particles (with sizes of approximately 100 nm), which may potentially exhibit a nanoporous structure. These inner geometrical properties are described by the pore size distribution. In addition to the pore size distribution, the inner geometry can be characterized using the specific surface area (SSA). In accordance with the Brunauer-Emmett-Teller (BET) analysis, this surface has the following expression: $A_{BET} = n_m a/M$, where $n_m$ is the monolayer capacity, $a$ is the adsorbed molecule cross-sectional area and $M$ is the mass of the adsorbent[15]. The pore size distribution is characterized by the properties of capillary condensation, which corresponds to adsorption/desorption hysteresis. In this case, we use one of the most standard adsorption analysis approaches, such as the Barrett-Joyner-Halenda (BJH) procedure and classical density functional theory (c-DFT) calculations[1]. Thus, the analysis of the adsorption isotherms provides desired information about the inner geometrical structure of synthesized materials.

**IX.** *Average particle size vs concentration, average particle size vs volume of load*

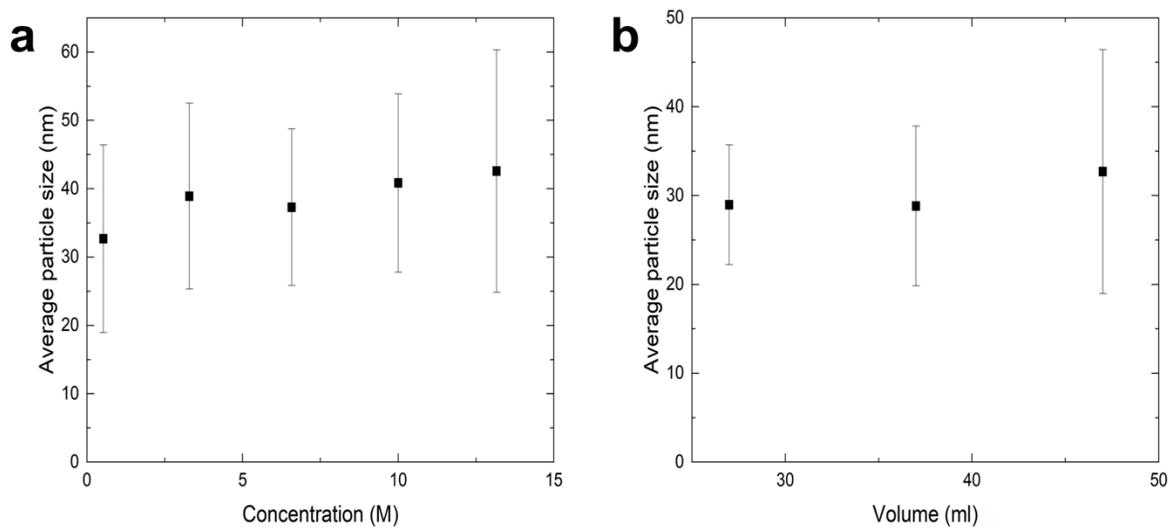

Figure S8: The dependence of the average particle size on (a), the concentration of ammonia (T = const = 423 K, m (Si) = 0.3 (b) filling of the autoclave.